\documentclass[fleqn,usenatbib]{mnras}

\usepackage{newtxtext,newtxmath}
\usepackage[T1]{fontenc}

\DeclareRobustCommand{\VAN}[3]{#2}
\let\VANthebibliography\thebibliography
\def\thebibliography{\DeclareRobustCommand{\VAN}[3]{##3}\VANthebibliography}

\usepackage{graphicx}	% Including figure files
\usepackage{amsmath}	% Advanced maths commands
\usepackage{subcaption}
\usepackage{xspace}
\newcommand{\dd}{{\rm d}}

\newcommand{\gyr}{\mbox{${\rm Gyr}$}}
\newcommand{\pc}{\mbox{${\rm pc}$}}
\newcommand{\mh}{\mbox{${\rm [M/H]}$}}
\newcommand{\K}{\textit{K}\xspace}

\newcommand{\hst}{\textit{HST}\xspace}
\newcommand{\jwst}{\textit{JWST}\xspace}
\newcommand*\code[1]{\textsc{#1}}

\title[$K$-corrections for star clusters]{RESCUER: Cosmological $K$-corrections for star clusters}

\author[M.~Reina-Campos \& W.~E.~Harris]{
Marta Reina-Campos$^{1,2}$\thanks{E-mail: reinacampos@mcmaster.ca (MRC)} 
and William E. Harris$^{1}$\thanks{E-mail: harrisw@mcmaster.ca (WEH)}
\\
$^{1}$Department of Physics \& Astronomy, McMaster University, 1280 Main Street West, Hamilton, L8S 4M1, Canada\\
$^{2}$Canadian Institute for Theoretical Astrophysics (CITA), University of Toronto, 60 St George St, Toronto, M5S 3H8, Canada\\
}

\date{Accepted XXX. Received YYY; in original form ZZZ}

\pubyear{2023}

\begin{document}
\label{firstpage}
\pagerange{\pageref{firstpage}--\pageref{lastpage}}
\maketitle

\begin{abstract}

The advent of \jwst (the \textit{James Webb Space Telescope}) now allows entire star cluster populations to be imaged in galaxies at cosmologically significant redshifts, bringing with it the need to apply \K-corrections to their magnitudes and colour indices. Since the stellar populations within star clusters can be well approximated by a single age and metallicity, their spectral energy distributions are very different from those of galaxies or supernovae, and their \K-corrections behave differently. We derive the photometric \K-corrections versus redshift for model star clusters that cover a wide range of ages and metallicities, illustrating the results particularly for the broadband filters on the \hst/ACS and the \jwst/NIRCam cameras that are most commonly being used for imaging of populations of star clusters in distant galaxies. In an Appendix, we introduce a simple webtool called RESCUER that can generate \K-values for any user-defined combination of cluster properties.

\end{abstract}


%%%%%%%%%%%%%%%%%%%%%%%%%%%%%%%%%%%%%%%%%%%%%%%%%%

%%%%%%%%%%%%%%%%% BODY OF PAPER %%%%%%%%%%%%%%%%%%

\section{Introduction}

The James Webb Space Telescope (\jwst) has opened up the ability to observe entire populations of star clusters at distances and lookback times well beyond the Local Universe \citep{faisst+2022,lee+2022,harris_reina-campos2023}.  But before the photometry of cosmologically distant systems can be compared with similar data for their zero-redshift counterparts, \K-corrections need to be applied to account for the effects of the cosmological redshift on the measured magnitudes and colour indices.  

\K-corrections in their various forms have long been familiar in the literature for galaxies and supernovae \citep[][to cite only a few]{hubble1936,humason+1956,oke_sandage1968,hamuy+1993,kim+1996,lubin_sandage2001,hogg02,blanton_roweis2007,boldt+2014}.  But they are almost unknown for photometry of star clusters \citep[see][for rare examples where globular clusters were observed in galaxies with significant redshifts]{kalirai+2008,alamo-martinez+2013,harris_reina-campos2023}.  The essential problem is that the SEDs (spectral energy distributions) for star clusters are not the same as those of galaxies, and they vary with redshift in a different way.  For the composite stellar populations that make up  galaxies, SED shapes (and thus $K-$corrections for any redshift) are determined by their morphological type, or more precisely their \emph{star formation history}.  By contrast, star clusters are close approximations to single-age SSPs (Simple Stellar Populations), and the major factors determining their SED shapes are instead \emph{metallicity} and \emph{age}.  A separate treatment of the problem specifically for star clusters is therefore appropriate, and timely for upcoming \jwst data.

In the following discussion, we adapt the general theory for $K-$corrections to SEDs of star clusters and demonstrate how $K-$values change with redshift up to $z=1$. We describe the formalism in Sect.~\ref{sec:kcorr}, and introduce model SEDs from the E-MILES stellar library with a selected set of filters in Sect.~\ref{sec:data}.  In Sect.~\ref{sec:results}, which has the main results of our paper, we start with a simple example for a blackbody spectrum, and then go on to calculate and discuss full \K-corrections for more realistic star clusters, as observed through selected filters for \jwst and \hst (\textit{Hubble Space Telescope}).  We briefly summarize our findings in Sect.~\ref{sec:summary}.

In this work, we use the cosmological parameters from \textit{Planck} 2018 \citep{planck2020}: $H_0=67.7~{\rm km/(Mpc\,s)}$, $\Omega_{\rm m}=0.31$, in their \emph{astropy} implementation.

\section{Cosmological \K-corrections}\label{sec:kcorr}

The cosmological photometric \K-correction quantifies the flux difference from a source at a redshift $z$ relative to its intrinsic luminosity. This correction is often needed in galaxy surveys to correct the observed apparent magnitudes into a uniform system of absolute magnitudes \citep[e.g.][]{blanton_roweis2007}, but it has not yet been calculated in a systematic way for star clusters. 

By convention, we define the \K-correction in terms of the absolute and apparent magnitudes \citep{hogg02,hogg2022},
\begin{equation}
    M_{\rm Q} = m_{\rm R} - 5\log_{10}\left(\dfrac{d_{\rm L}}{10~\pc}\right) - K_{\rm QR}, \label{eq:k-definition} 
\end{equation}
where $M_{\rm Q}$ is the rest-frame absolute magnitude of the source (i.e.~the magnitude if the source were to be observed at $10~\pc$) in the filter $Q$, and $m_{\rm R}$ the apparent magnitude of the source observed in the (possibly different) filter $R$. The luminosity distance $d_{\rm L}$ to the source depends on the cosmology assumed, and in a flat Universe, it is proportional to the comoving distance to the source, $d_{\rm L} = (1+z)d_{\rm C}$ \citep[e.g.][]{hogg1999}.

The \K-correction can be defined in terms of either frequency $\nu$ or wavelength $\lambda$ \citep{hogg02}, but in the present discussion, only the more common wavelength version is presented, and only in terms of photon-counting instruments.
As will be seen below, the calculation of \K is built from various integrals of the general form
\begin{equation}
    N = \int \frac{f(\lambda)}{hc/\lambda} T(\lambda) d\lambda = \frac{1}{hc} \int \lambda f(\lambda) T(\lambda) d\lambda \, .
\end{equation}
Here $N$ is the number of recorded photons per unit time per unit area, from a source with flux $f(\lambda)$, measured by a detector with overall throughput (i.e.~transmission profile) $T(\lambda)$ for a given filter.  The flux $f$ describing the SED of the source is assumed to be in units of energy/time/area/wavelength, so it is converted to counts/time/area/wavelength by dividing by the energy per photon $(hc/\lambda)$. The transmission profile $T$ represents the entire system throughput, which for \jwst and \hst includes the telescope optics, camera, filter, and detector efficiencies as a function of wavelength\footnote{For ground-based instruments, the atmospheric transmission would also be included.}. Thus 
$T(\lambda)$ essentially gives the probability that an incoming photon of wavelength $\lambda$ will be recorded by the detector. 

In terms of wavelength, the general expression for $K_{QR}$ is \citep[see][for derivation]{hogg02}
%Expanding on the definitions of the absolute and apparent magnitudes, the \K-correction can be calculated in terms of the spectral flux density of the source $f_{\lambda}$ \citep[i.e.~the energy per unit time, per unit wavelength and per unit area; see][for a detailed derivation]{hogg02},
\begin{equation}
\begin{aligned}
    K_{\rm QR} &= -2.5\log_{10}\left(\dfrac{1}{(1+z)}\times \right. \\ &\left.\dfrac{\int \dd \lambda \lambda_{\rm o} f_{\lambda}(\lambda_{\rm o})R(\lambda_{\rm o}) \int \dd \lambda \lambda_{\rm rf} g_{\lambda}^{Q}(\lambda_{\rm rf})Q(\lambda_{\rm rf})}{\int \dd \lambda \lambda_{\rm o} g_{\lambda}^{R}(\lambda_{\rm o})R(\lambda_{\rm o}) \int \dd \lambda \lambda_{\rm rf} f_{\lambda}[(1+z)\lambda_{\rm rf}]Q(\lambda_{\rm rf})}\right).\label{eq:kcorr-flux}
\end{aligned}
\end{equation} 

The integrals cover the range of observed and rest-frame wavelengths, $\lambda_{\rm o}$ and $\lambda_{\rm rf}$\footnote{Note that previous studies write the rest-frame wavelength as $\lambda_e$ or `emitted'.  We prefer to use $\lambda_{\rm rf}$ to make the distinction with $\lambda_o$ more general; see the discussion below.}, respectively, and the terms $R(\lambda)$ and $Q(\lambda)$ describe the transmission curves of the two filters. %These bandpasses correspond to how many photons per unit wavelength the detector will count in the wavelength range of a given filter. 
For these we use the published throughput curves for \jwst/NIRCam\footnote{\href{https://jwst-docs.stsci.edu/jwst-near-infrared-camera/nircam-instrumentation/nircam-filters}{https://jwst-docs.stsci.edu/jwst-near-infrared-camera/nircam-instrumentation/nircam-filters}} and \hst/ACS\footnote{\href{https://www.stsci.edu/hst/instrumentation/acs/data-analysis/system-throughputs}{https://www.stsci.edu/hst/instrumentation/acs/data-analysis/system-throughputs}}.
The terms $g_{\lambda}^{Q}$ and $g_{\lambda}^{R}$ correspond to the spectral flux densities of the standard source in the filters $Q$ and $R$, respectively. In this work we use the AB magnitude system \citep{oke_gunn1983}, where the magnitudes are defined in terms of a hypothetical constant source of flux density in frequency space, $g_{\nu}^{\rm AB} \equiv 3.631 \times 10^{-20} {\rm erg}~{\rm cm}^{-2}~{\rm s}^{-1}~{\rm Hz}^{-1}$ at all frequencies. The spectral density of this source can be transformed to wavelength space by $g_{\lambda} = g_{\nu}(c/\lambda^2)$, using $\nu g_{\nu} = \lambda g_{\lambda}$ and 
%the relation between the speed of light, wavelength and frequency, 
$c = \lambda\nu$.

The \K-correction can also be described in terms of the intrinsic luminosity of the source, $L_{\lambda}(\lambda)$ (i.e.~the energy per unit time per unit wavelength). This expression can be related to its spectral flux density via the luminosity distance and redshift, $L_{\lambda} (\lambda_{\rm o}) = (1+z) 4\pi d_{\rm L}^2 f_{\lambda}(\lambda_{\rm rf})$. Replacing all terms in eq.~(\ref{eq:kcorr-flux}), the \K-correction is thus
\begin{equation}
\begin{aligned}
    K_{\rm QR} &= -2.5\log_{10}\left(\dfrac{1}{(1+z)}\times \right. \\ &\left.\dfrac{\int \dd \lambda \lambda_{\rm o} L_{\lambda}\left[(1+z)^{-1}\lambda_{\rm o}\right]R(\lambda_{\rm o}) \int \dd \lambda \lambda_{\rm rf} g_{\lambda}^{Q}(\lambda_{\rm rf})Q(\lambda_{\rm rf})}{\int \dd \lambda \lambda_{\rm o} g_{\lambda}^{R}(\lambda_{\rm o})R(\lambda_{\rm o}) \int \dd \lambda \lambda_{\rm rf} L_{\lambda}(\lambda_{\rm rf})Q(\lambda_{\rm rf})}\right). \label{eq:kcorr-luminosity}
\end{aligned}
\end{equation}

The \K-values written this way respond to the question of \textit{how much the observed apparent magnitude should be corrected in order to reflect the intrinsic luminosity}. 
%This general way of defining the \K-correction allows for the observed filter $R$ and the emitted one $Q$ to be different, and thus, 
The calculation of \K~  can go in two different directions that we refer to as \emph{homochromatic} and \emph{heterochromatic}:

\begin{enumerate}
    \item \emph{Homochromatic} ($Q = R$): The correction is done within the same wavelength or frequency range as the observed (redshifted) measurement. For a homochromatic \K-correction in the filter $Q$, the previous equations simplify to 
\begin{equation}
\begin{aligned}
    K_{\rm QR} &= -2.5\log_{10}\left(\dfrac{1}{(1+z)} \dfrac{\int \dd \lambda \lambda f_{\lambda}\left(\lambda\right)Q(\lambda)}{ \int \dd \lambda \lambda f_{\lambda}\left[(1+z)^{-1}\lambda\right]Q(\lambda)}\right)\\
    &=-2.5\log_{10}\left(\dfrac{1}{(1+z)} \dfrac{\int \dd \lambda \lambda L_{\lambda}\left[(1+z)^{-1}\lambda\right]Q(\lambda)}{ \int \dd \lambda \lambda L_{\lambda}(\lambda)Q(\lambda)}\right).
\end{aligned}
\end{equation}
This version is by far the most frequently used one and intuitively clear:  the observed flux at wavelength $\lambda$ was emitted in the rest-frame spectrum at the shorter wavelength $\lambda/(1+z)$, and the energy of every photon is reduced by the same factor $1/(1+z)$. 
    \item \emph{Heterochromatic} ($Q \neq R$):  This more general formulation allows the transformation from the observed wavelength $\lambda_{\rm o}$ to be done to another wavelength $\lambda_{\rm rf}$ on the rest-frame (emitted, un-redshifted) spectrum.  For example, an obvious use of a heterochromatic conversion would be where filter $R$ is just the redshifted version of filter $Q$ in which the flux was emitted, such that $\lambda_{\rm rf} = \lambda_{\rm e} = \lambda_{\rm o}/(1+z)$. 

    However, Eq.~(\ref{eq:kcorr-luminosity}) is general enough to  allow the transformation to be done into any arbitrary filter where the rest-frame $\lambda_{\rm rf}$ does not have to equal $\lambda_o/(1+z)$.  If 
    %we measure the magnitude from the redshifted SED through filter $R$ covering the wavelength range $\Delta \lambda$, and 
    we are given an accurate SED, then the rest-frame magnitude through \emph{any} other filter $Q$ at \emph{either} shorter or longer wavelength can be predicted from the measured flux through $R$ \citep[e.g.][]{kim+1996,blanton_roweis2007}. 
    %As noted above, some studies focus just on transforming the observed flux back to the wavelength range at which it was emitted, %i.e.~ $(1+z)^{-1}\Delta \lambda$ centered at 
    %$\lambda_{\rm rf} = \lambda_{\rm e} = \lambda_{\rm o}/(1+z)$ \red{REF}.  But 
     This is the main reason why we use the term $\lambda_{\rm rf}$ and not $\lambda_{\rm e}$ to refer to the rest-frame wavelength.

    In short, the \K-correction can in principle be used to step from the observed flux at $\lambda_o$ to any point on the rest-frame spectrum, but clearly the validity of the result will depend heavily on the accuracy of the assumed SED.  In this sense, working with SEDs for star clusters, which closely resemble blackbody-like SSPs (see below), is more straightforward than for the far more complex parameter space needed to model galaxy spectra \citep[cf.][for an extensive discussion of galaxy template spectra]{blanton_roweis2007}.  
 \end{enumerate}

\begin{figure*}
	\includegraphics[width=\hsize]{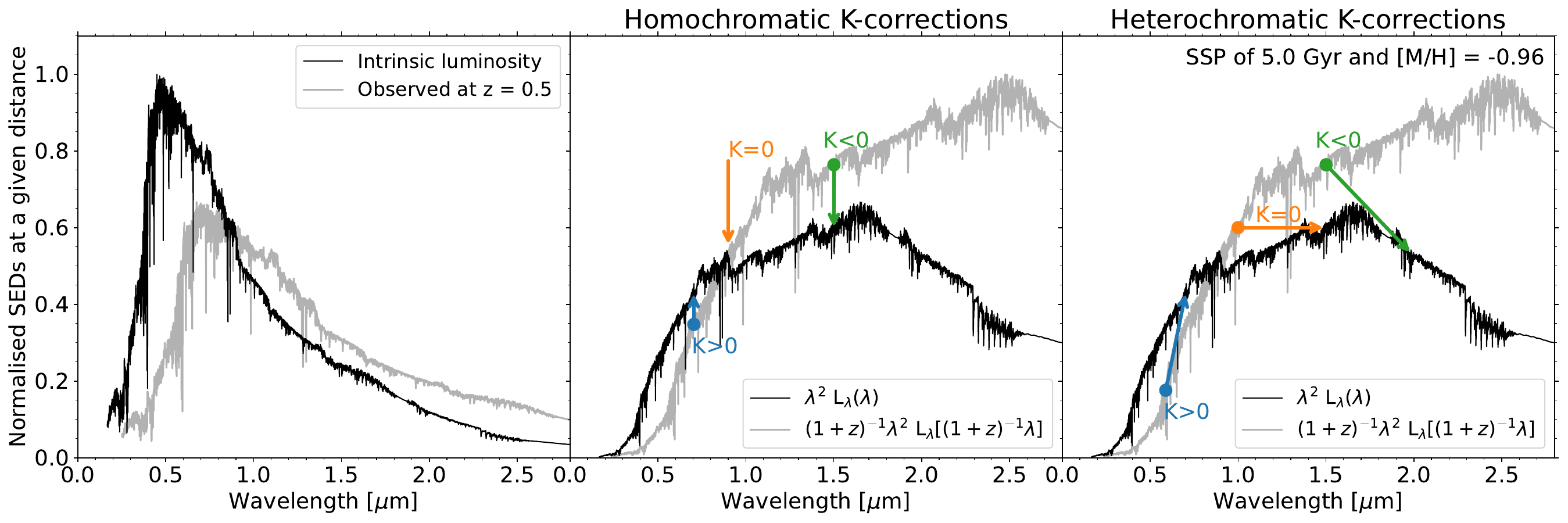}
    \caption{Expected behaviour of the \K-correction on the SED of a SSP of $5~\gyr$ and $\mh=-0.96$ observed at $z=0.5$: (\textit{left panel}) intrinsic luminosity emitted by the SSP and the observed SED at $z=0.5$,  (\textit{middle panel}) homochromatic corrections within the same wavelength range, (\textit{right panel}) heterochromatic corrections across different ranges of the spectrum. The curves in the left-hand panel are shown in units of $2.9\times10^{29}$ ergs s$^{-1}$ \AA$^{-1}$ M$_{\odot}^{-1}$, and the curves in the middle and right-hand panels are normalised by $2.2\times10^{37}$ \AA~ergs s$^{-1}$ M$_{\odot}^{-1}$. Positive \K-values correspond to the observed flux being dimmer than the intrinsic one (blue arrows) and the absolute magnitude needing to increase, whereas negative \K-values indicate that the observed flux is brighter (green arrows), and thus the magnitude has to decrease.} \label{fig:kcorr_pedagogical}
\end{figure*}

A simple way to build some further intuition is to assume that the filter transmission curves are described by delta functions, $Q(\lambda) = \delta(\lambda - \lambda_{\rm o})$; this is the \emph{monochromatic} version of \K as used in, e.g., \citet{condon_matthews2018}.
%i.e. that there is $100~$per cent transmission at a given wavelength and none for the rest of the spectrum. 
With this simplifying assumption, eq.~(\ref{eq:kcorr-luminosity}) reduces to
\begin{equation}
\begin{aligned}
    K_{\rm QR} = -2.5\log_{10}\left[\dfrac{1}{(1+z)} \dfrac{L_{\lambda}\left[(1+z)^{-1}\lambda_{\rm o}\right]}{L_{\lambda}(\lambda_{\rm rf})} \left(\dfrac{\lambda_{\rm o}}{\lambda_{\rm rf}}\right)^2\right].
\end{aligned}
\end{equation}
This equation is only valid in the AB magnitude system because its spectral flux density in frequency space is constant. In the case of homochromatic \K-corrections ($\lambda_{\rm o} = \lambda_{\rm rf}$), we recover the expression provided by \citet{condon_matthews2018} in their equation (67). 

The basic behaviour of the \K-correction is illustrated 
in Fig.~\ref{fig:kcorr_pedagogical}, where an intrinsic emitted SED for a simple stellar population (SSP) of 5 Gyr age and $\mh=-0.96$ is compared with its redshifted version at $z=0.5$ (left-hand panel). To show the intrinsic and emitted SEDs on the same scale, we assume for display purposes that the intrinsic curve (black line) represents the object at its observed distance $d_L$, but  sitting at rest, while the observed redshifted SED (shaded line) is attenuated and displaced towards redder wavelengths. This way we isolate just the effect of redshift.  

The middle and right-hand panels show the same pair of spectra now multiplied by $\lambda^2$.
Examples of homochromatic \K-corrections are shown in the middle panel, while heterochromatic \K-corrections are in the right-hand panel. The resulting sign of the \K-correction contains information about the shape of the SEDs. A positive (\K$>0$) value (blue arrows) indicates that the observed flux is dimmer than the intrinsic flux at that wavelength, and thus the absolute magnitude should be brighter than expected from the inverse-square law (Eq.~\ref{eq:k-definition}). In contrast, a negative (\K$<0$) value (green arrows) makes the absolute magnitude fainter because the observed flux is brighter than the intrinsic flux.
%\footnote{Note that the \K-correction is defined with a negative sign in front in eq.~(\ref{eq:k-definition}).}. 
When both curves are equal, the \K-correction is null (orange arrows). 
%Because of the blackbody-like shape of their SEDs, redshifted star clusters look brighter in the infrared than expected from just the inverse-square law (\K$<0$), whereas they are much fainter in the optical/blue range (\K$>0$). 

\section{Data}\label{sec:data}

\subsection{Stellar population synthesis models: E-MILES}

\begin{figure*}
	\includegraphics[width=0.9\hsize]{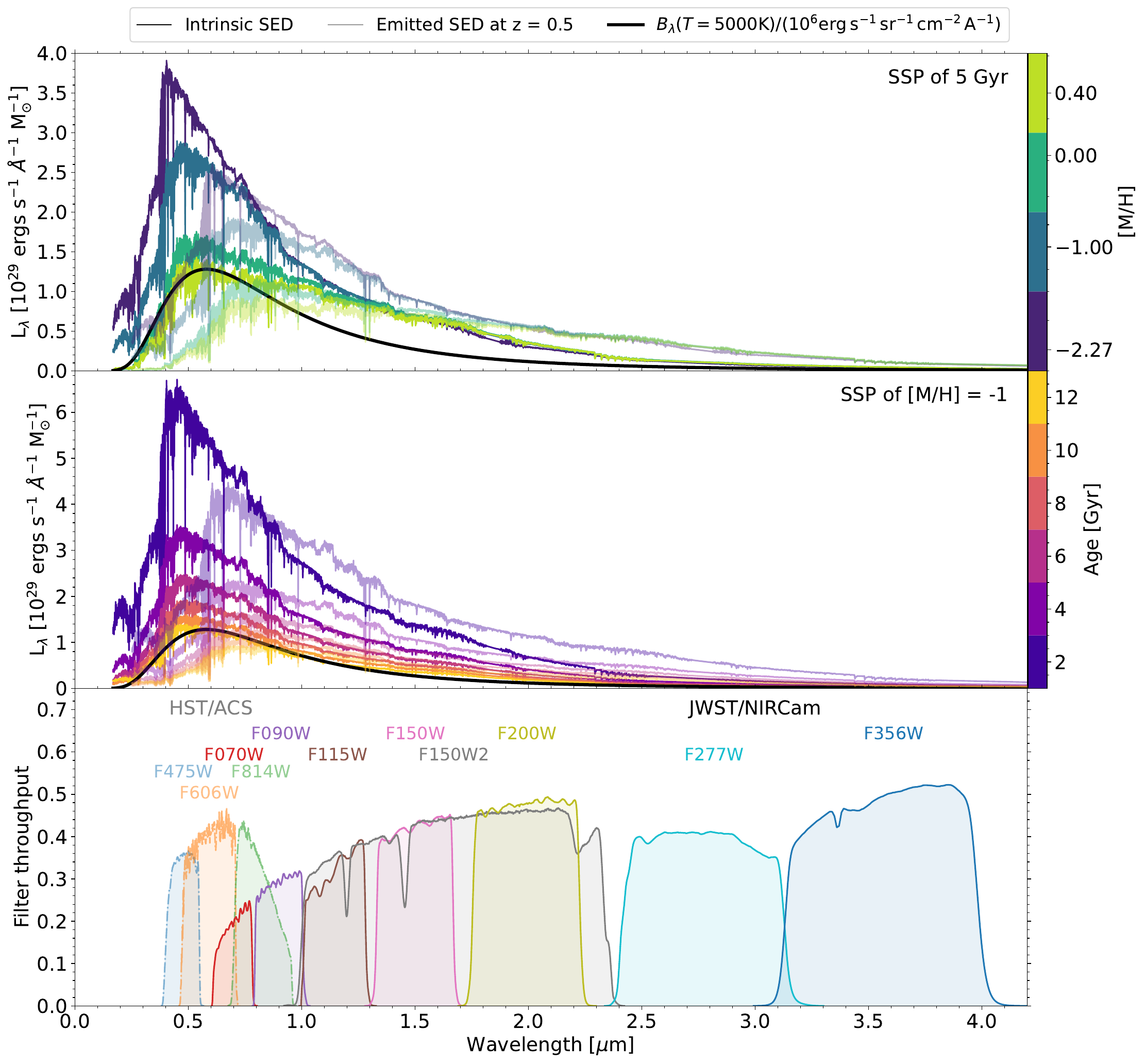}
    \caption{Luminosity of the E-MILES stellar population models for SSPs of different metallicities (\textit{top panel}) and ages (\textit{middle panel}). These models assume a \citealt{chabrier2003} IMF and the BaSTI isochrones from \citet{pietrinferni2004}. Solid lines correspond to the intrinsic luminosities, and the transparent lines show the SED emitted if the SSP where to be located at $z=0.5$. The black solid line corresponds to the blackbody spectrum at $T=5000~$K with an arbitrary normalization. The bottom panel shows the bandpasses of a variety of filters from the \hst/ACS and the \jwst/NIRCam cameras. } \label{fig:seds_mh_ages}
\end{figure*}

To make the calculations for \K, we need to work from a library of homogeneous SEDs that cover suitably large ranges in wavelength, metallicity, and age. For this study, we use SSPs calculated with the E-MILES models \citep{rock+2016}\footnote{The E-MILES stellar population models are publicly available here: \href{http://research.iac.es/proyecto/miles/pages/spectral-energy-distributions-seds/e-miles.php}{http://research.iac.es/proyecto/miles/pages/spectral-energy-distributions-seds/e-miles.php}}. The SEDs produced from these models cover the range $1680~$\AA--$50000~$\AA\, at high resolution, can be generated for any desired age or metallicity spanning the observed ranges for star clusters, and are well tested against observed SEDs for stellar systems \citep[e.g.][]{rock+2016,vazdekis+2015,vazdekis+2016}. More generally, several modern SSP codes are now available that accurately match the integrated spectra of real globular clusters in the Milky Way or M31 from the UV through the infrared \citep[e.g.][among others]{barber+2014,conroy+2018,ashok+2021,martins+2019,boquien+2019,maraston+2020} and several of these would be similarly useful for the purposes of this study.

%Some references on SSP models for globular cluster spectra:  the main message is that modern SSP codes are well tested against integrated spectra of GCs (usually the Milky Way or M31) and reliable.  \citet{conroy+2018} compare their SSP models with Schiavon Milky Way GC spectra and show good matches.  \citet{vazdekis+2015,vazdekis+2016} match MILES and E-MILES SSPs with GCs, emphasizing the UV region.  \citet{ashok+2021} present A-LIST grid built from MIST and PARSEC isochrones, and tested against M31 GC spectra.  \citet{martins+2019} test various SSPs including MILES against optical spectra of Milky Way GCs, and \citet{barber+2014} do something similar.  Some of these papers discuss the spectral resolution of the model SEDs.  Not sure this is important for us -- K-corrections are all about the redshift effects on \emph{broadband} filters, so spectral resolution probably doesn't matter.

The E-MILES stellar population models can be generated for different choices of stellar initial mass functions (IMF) and theoretical stellar isochrones. In this work, we use the models derived in version v11.0 assuming a \citealt{chabrier2003} IMF and the BaSTI isochrones from \citet{pietrinferni2004}. In Appendix~\ref{app:diff-seds}, we show that the results presented here are only mildly affected ($\sim0.02~\rm mags$ by $z = 0.5$, and $\sim0.12~\rm mags$ by $z = 1$) when assuming Padova isochrones instead \citep{girardi2000}, and remain unchanged for a \citet{kroupa2001} IMF.

An illustration of the SEDs from the E-MILES models is shown in Fig.~\ref{fig:seds_mh_ages}. Regardless of the age or metallicity of the SSP, the shape of the SEDs is remarkably similar to that of a blackbody at $T=5000~$K; the SEDs are dominated by a single peak in the optical with a decay towards redder wavelengths. Comparing the SEDs of $5~\gyr$ old SSPs with different metallicities (top panel), the peaks of those with lower metallicity are more prominent by a factor of $2.9$ than in those with super-solar abundances. In contrast, for a given metallicity, young SSPs ($\tau = 2~\gyr$) emit more radiation across their spectrum than old SSPs ($\tau = 12~\gyr$).

\subsection{Filter selection: \hst and \jwst}

For the selection of filters, we focus on three commonly used \hst filters for GC systems in other galaxies ($F475W$, $F606W$ and $F814W$), plus a set of eight broadband filters for \jwst NIRCam. These include the SWC (short wavelength channel) filters ($F070W$, $F090W$, $F115W$, $F150W$, $F150W2$, and $F200W$) and the LWC (long wavelength channel) ($F277W$ and $F356W$). We show their bandpasses in the bottom panel of Fig.~\ref{fig:seds_mh_ages}. These NIRCam filters have already been used for GC photometry in high-$z$ systems \citep{faisst+2022,lee+2022,harris_reina-campos2023} and more studies are in progress.

It is worth noting that the formulation presented here for the cosmological \K-corrections is general and easily adaptable to any other set of filters for which the bandpass is known. The version of the webtool RESCUER\footnote{\href{https://rescuer.streamlit.app}{https://rescuer.streamlit.app}} presented below in Appendix \ref{app:webtool} is restricted to this set of \hst and \jwst filters, but the code is easily adaptable via the public repository on GitHub.

\section{Results}\label{sec:results}

The \K-corrections presented in this work are in the AB magnitude system \citep{oke_gunn1983}.

\subsection{Test case: blackbody spectrum}

The SEDs of star clusters over the age range of interest here are approximated well to first order by Planck blackbody spectra (see Fig.~\ref{fig:seds_mh_ages}), which can be used to give an initial impression of the behaviour of \K~ versus redshift for the set of filters listed above.
Consider the spectrum emitted by a blackbody at $T=5000~$K. 
The spectral radiance of the blackbody, i.e.~the energy per unit time, per unit solid angle, and per unit of area normal to the propagation, in terms of wavelength is
\begin{equation}
    B_{\lambda}(\lambda, T) = \dfrac{2hc^2}{\lambda^5}\left[\exp\left(\dfrac{hc}{\lambda k_{\rm B} T}\right)-1\right]^{-1}
\end{equation}
where $h$ is the Planck constant, and $k_{\rm B}$ is the Boltzmann constant.

\begin{figure}
	\includegraphics[width=\hsize]{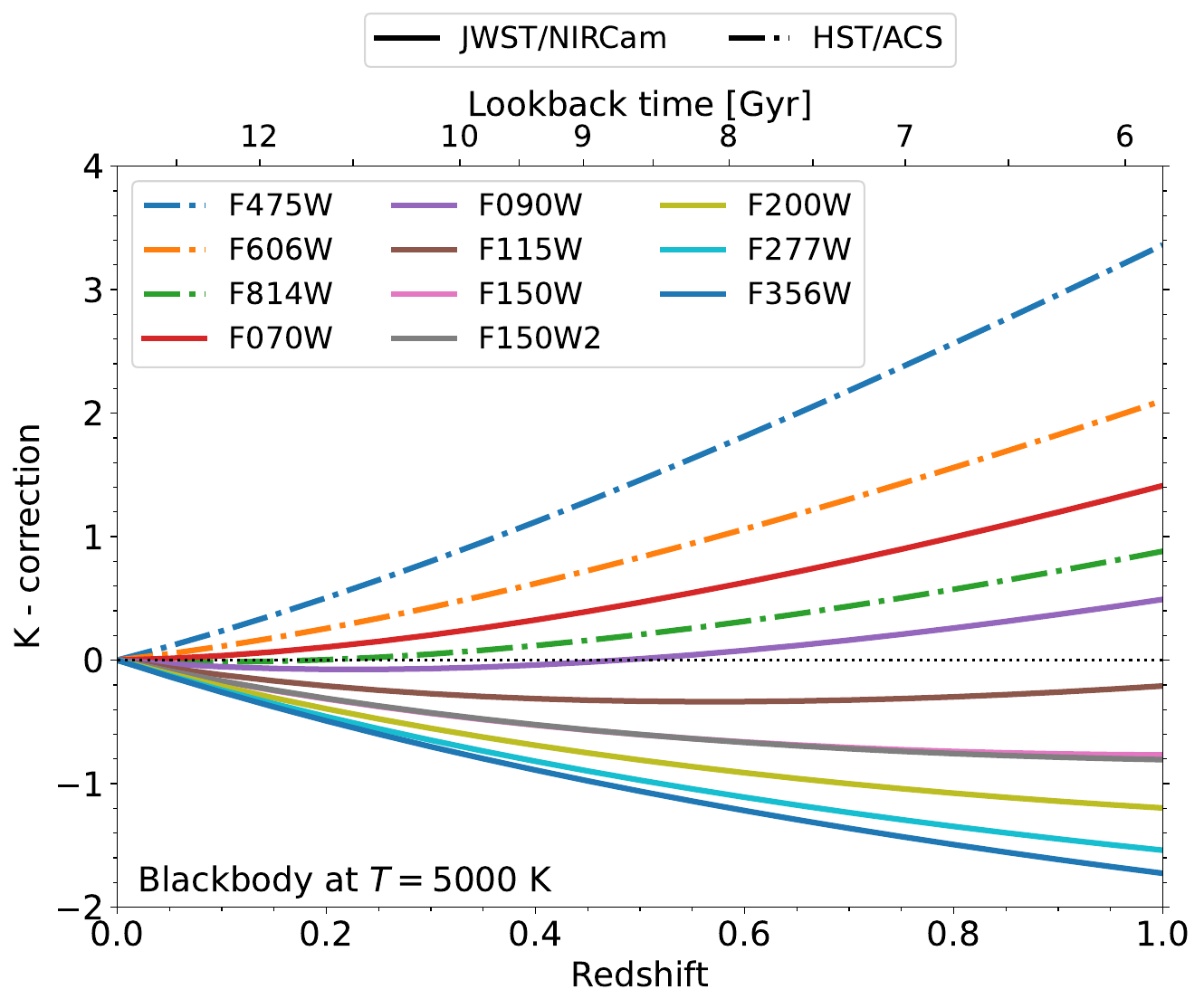}
    \caption{Homochromatic K-corrections to be applied to the magnitudes of a blackbody at $T=5000~$K calculated for different filters as a function of redshift of the source. Solid lines correspond to the \jwst filters, and dotted-dash lines show the three \hst filters.} \label{fig:kcorr_blackbody}
\end{figure}

We calculate the homochromatic \K-corrections 
%that would need to be applied to such spectra as a function of redshift ($z\leq 1$) 
using Eq.~(\ref{eq:kcorr-flux}), and show them in Fig.~\ref{fig:kcorr_blackbody}. As a sanity check, all filters require a null \K-correction at $z=0$. At higher $z$, the value of the \K-correction for most filters can increase up to a few magnitudes. The filters for which the correction would be the smallest at all redshifts are $F090W$ and $F115W$. For both of these filters, the emission mostly comes from the region in the spectrum around the peak of the blackbody emission, where the curve is shallow. The peak of the blackbody emission crosses the wavelength range of $F090W$ at $z\sim0.5$, hence the sign change of the \K-correction from negative to positive. In the case of the filter $F115W$, the peak would cross it at $z>1$, and the sign change is thus not visible in the figure.

\subsection{The oldest SSPs of [M/H] = -2.27 as a function of redshift}

\begin{figure}
	\includegraphics[width=\hsize]{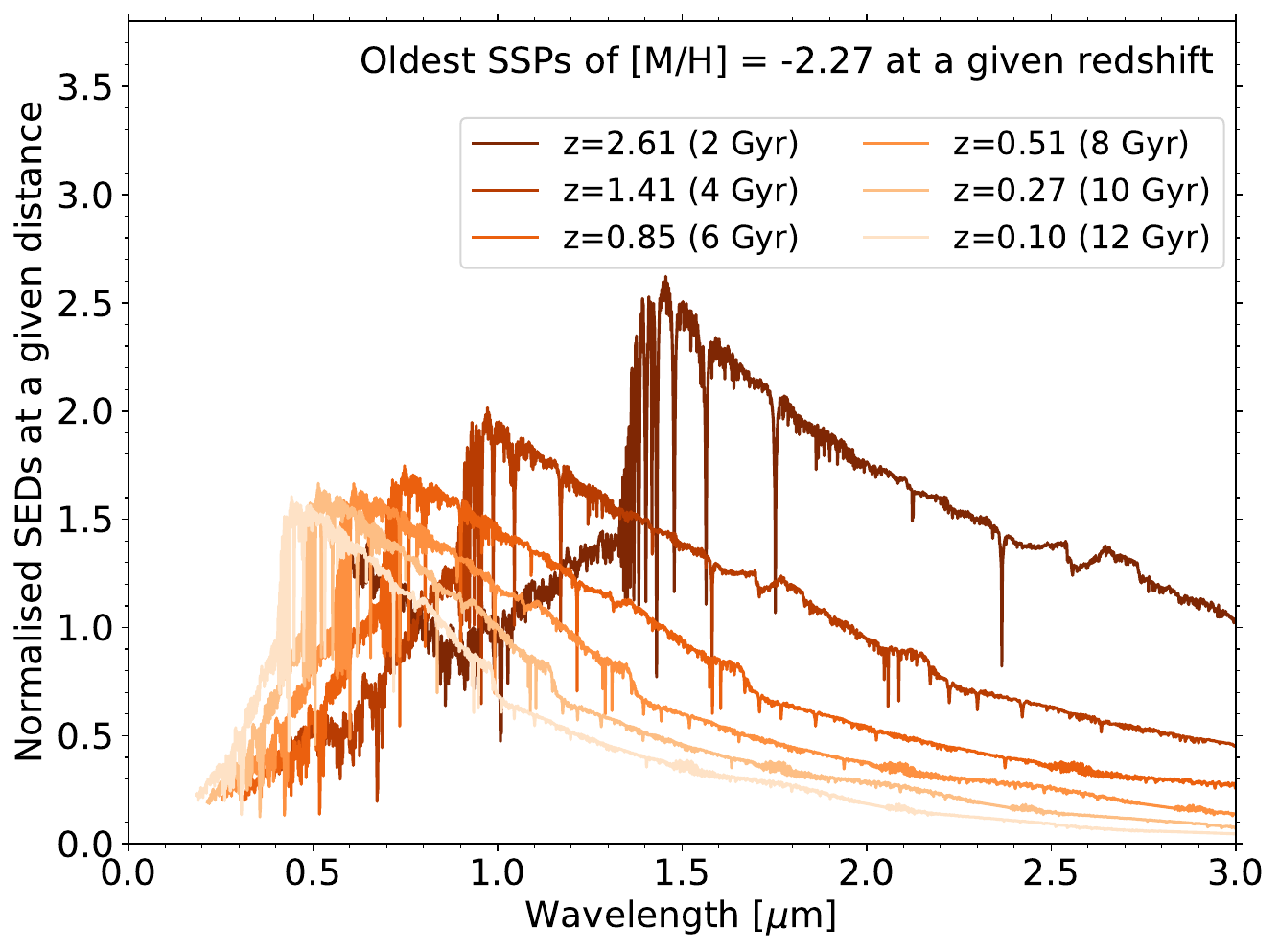}
 	\includegraphics[width=\hsize]{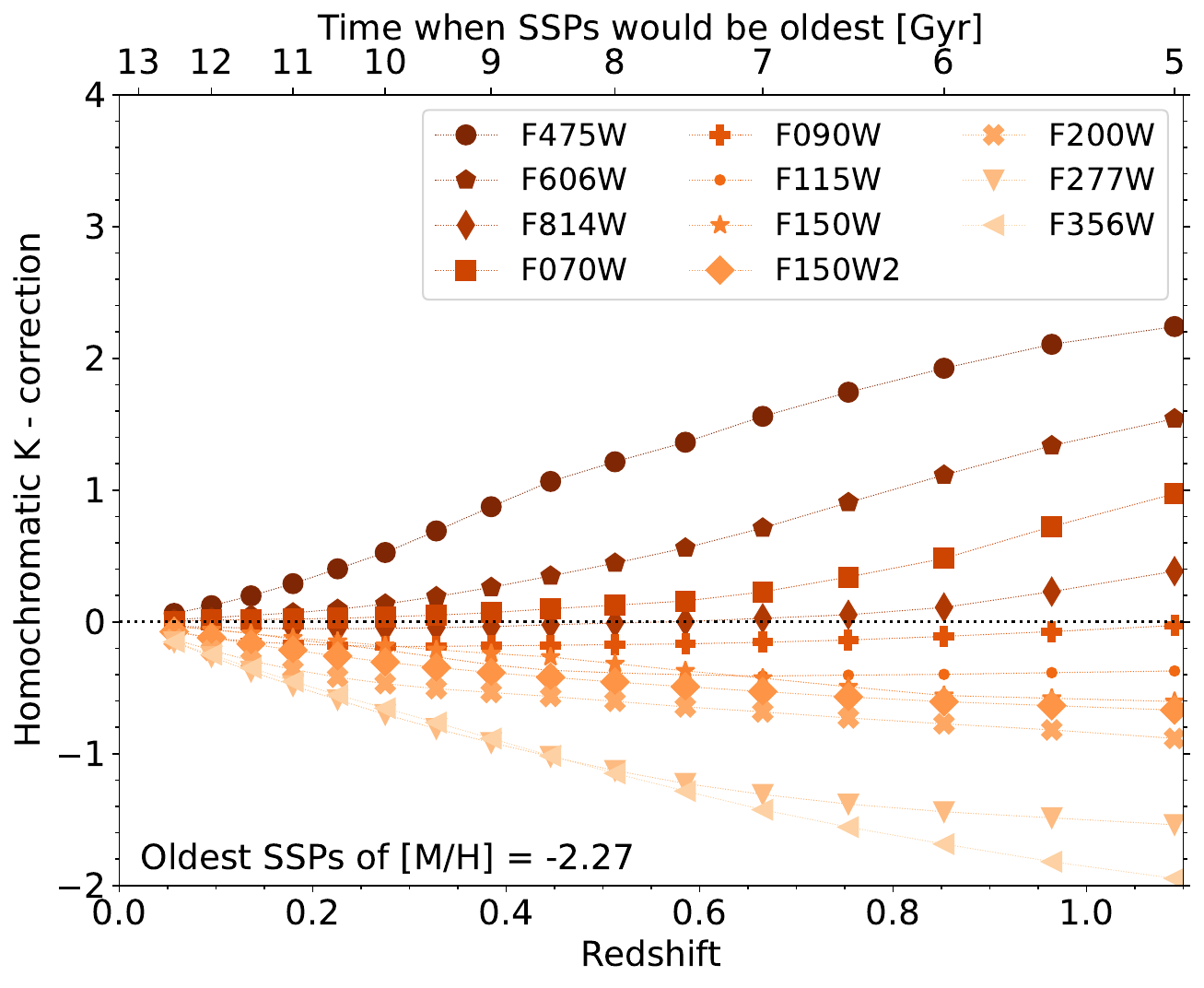}
    \caption{The oldest SSPs as a function of redshift. (\textit{Top}): expected attenuated SEDs of the oldest SSPs with $\mh=-2.27$ at different redshifts assuming that the sources are at the same distance. The SEDs are shown in units of $10^{29}$ ergs s$^{-1}$ \AA$^{-1}$ M$_{\odot}^{-1}$. (\textit{Bottom}): homochromatic K-corrections to be applied to the oldest metal-poor SSPs as a function of redshift calculated for different filters.} \label{fig:kcorr_oldest}
\end{figure}

An extremely interesting consequence of observing remote star clusters is that they have a well defined maximum observable age at a given redshift: that is, $t_{\rm max}(z) = (t_{\rm BB}(z) - t_{\rm form})$, where $t_{\rm BB}$ is the time since the Big Bang and $t_{\rm form}$ is the time interval needed for galaxy and star cluster formation to start.  Drawing from recent observations \citep[e.g.][]{labbe2023}, we adopt $t_{\rm form} \simeq 500$ Myr for the present discussion.
%is that we are observing them at preliminary stages of their evolution. Thus, for a given metallicity, we can explore the SEDs of proto-globular clusters, i.e.~the oldest SSPs at a given redshift (shown in the top panel of Fig.~\ref{fig:kcorr_oldest}).  These stellar populations can be understood as the progenitors of the old ($\tau>12~\gyr$) globular clusters in the local Universe. 
%For this, we assume that SSPs take $\sim 500~\myr$ to start forming after the Big Bang \citep{labbe2023}. 
Under this assumption, the stellar population of a $\sim13~\gyr$ old globular cluster at $z=0$ would (for example) have been $8~\gyr$ old at $z=0.51$, and $2~\gyr$ old at $z=2.61$. Despite the attenuation introduced by redshift, the displacement of the peak into redder wavelengths and the increase in luminosity from the younger populations (see top panel in Fig.~\ref{fig:kcorr_oldest}) implies that high-$z$ star clusters remain bright in the infrared and should be detectable by \jwst up to $z\sim 1$ and possibly beyond, even without the help of lensing. 

We now calculate the homochromatic \K-corrections for the model SEDs of age $t_{\rm max}(z)$  versus $z$, and show them in the bottom panel of Fig.~\ref{fig:kcorr_oldest}. By redshift $z=1$, the required \K-corrections range from  $-2$ to $2~$mags, and there are three filters ($F814W$, $F090W$ and $F115W$) for which they stay within $[-0.5, 0.5]~\rm mags$. As in the case of the blackbody spectrum, these three filters mostly capture emission coming from around the peak of the spectrum where it is shallower. Due to the number of uncertainties in the stellar population models, large \K-values might be introducing large errors into the magnitudes, and therefore, filter transformations involving smaller \K-corrections should be preferred. We have repeated the analysis for SSPs of metallicities $\mh=-0.96$ and $\mh = 0.06$ and the conclusions do not change. 

\begin{figure}
	\includegraphics[width=\hsize]{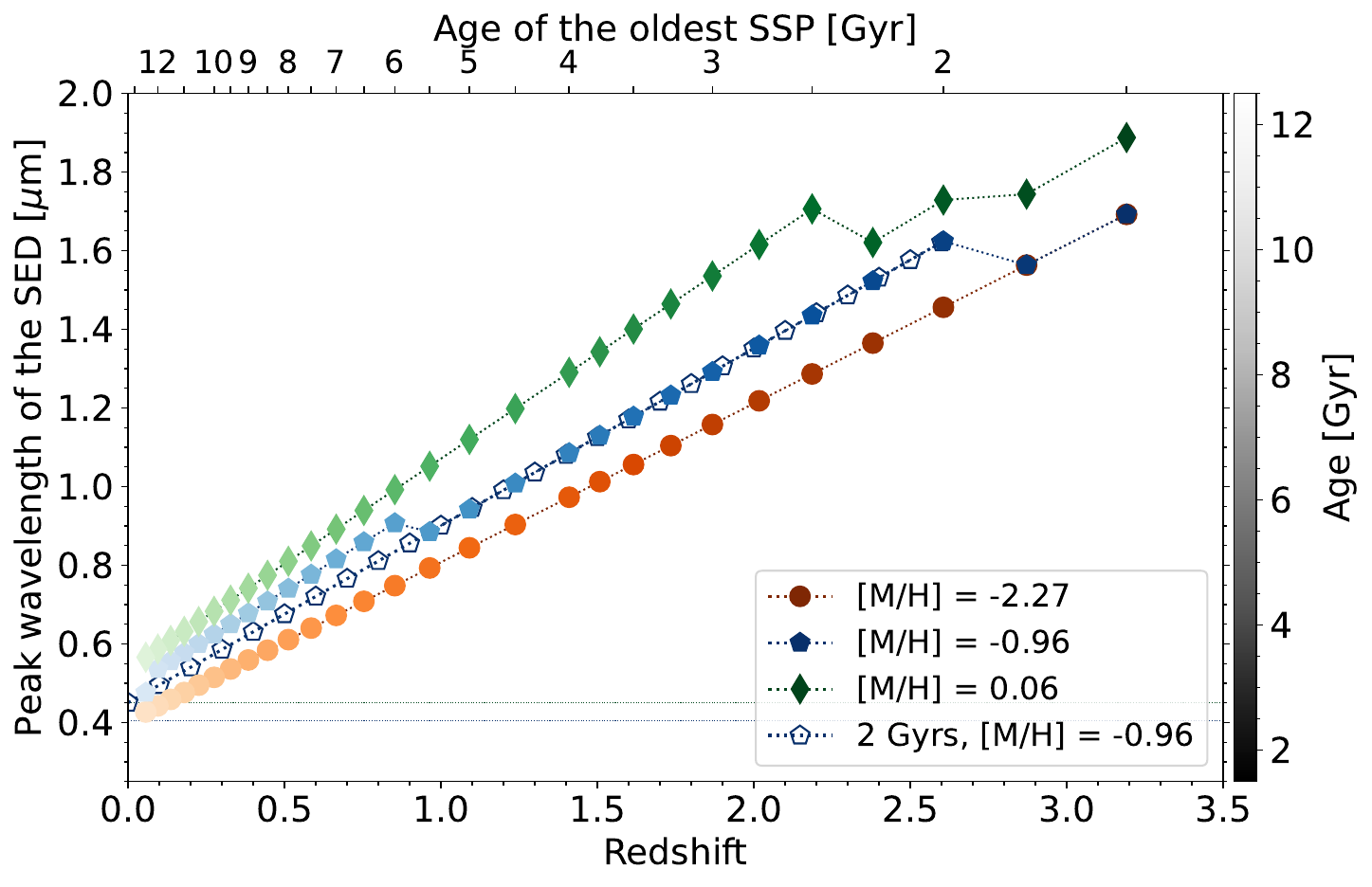}
    \caption{Wavelength at which the SED peaks as a function of redshift for the oldest SSPs at that redshift. Different color schemes correspond to three metallicities as indicated in the legend, and the shading of the marker shows the age of the SSP. The horizontal dotted lines correspond to the peaks of the intrinsic luminosity curves, and the empty markers indicate the redshift evolution of the peak of the SED of a SSP of $2~\gyr$ and $\mh=-0.96$.} \label{fig:seds_peak}
\end{figure}

As seen in the top panel of Fig.~\ref{fig:kcorr_oldest}, there are two competing factors on the displacement of observed SED: as redshift increases, the observed wavelengths get stretched and redder, but conversely the populations of the clusters are younger and their intrinsic SED is bluer. To explore these competing effects, we determine the wavelength at which the SED peaks as a function of redshift for a variety of SSPs of different ages and metallicities (Fig.~\ref{fig:seds_peak}). Because the peaks of the intrinsic SEDs for a given metallicity are roughly the same (see top and middle panel in Fig.~\ref{fig:seds_mh_ages}), then the displacement is mostly given by the stretching due to redshift. Interestingly, the peaks of the SEDs do not enter the infrared regime until $z>1$. This implies that one of the main advantages of \jwst over \hst for distant populations of star clusters is its higher resolution and much larger collecting area, rather than its infrared capability.  The advantage gained from imaging in the near-infrared comes from the \K-correction itself.
%In contrast, observations in the infrared region of the spectrum become crucial for high-$z$ studies of star clusters.

%\begin{figure*}
%	\includegraphics[width=\hsize]{imgs/sed_kcorr_z_attenuation_oldest_threemh.pdf}
% 	\includegraphics[width=\hsize]{imgs/ssp_oldest_ssps_Zp0.06_filters_threemh.pdf}
%    \caption{The oldest SSPs as a function of redshift. (\textit{Top}): expected attenuated SEDs of the oldest SSPs with $\mh=-2.27$ at different redshifts assuming that the sources are at the same distance. The SEDs are shown in units of $10^{29}$ ergs s$^{-1}$ \AA$^{-1}$ M$_{\odot}^{-1}$. (\textit{Bottom}): homochromatic K-corrections to be applied to the oldest metal-poor SSPs as a function of redshift calculated for different filters.} \label{fig:kcorr_oldest}
%\end{figure*}

\subsection{Homochromatic \K-corrections}\label{sub:homochromatic-ks}

\begin{figure*}
	\includegraphics[width=\hsize]{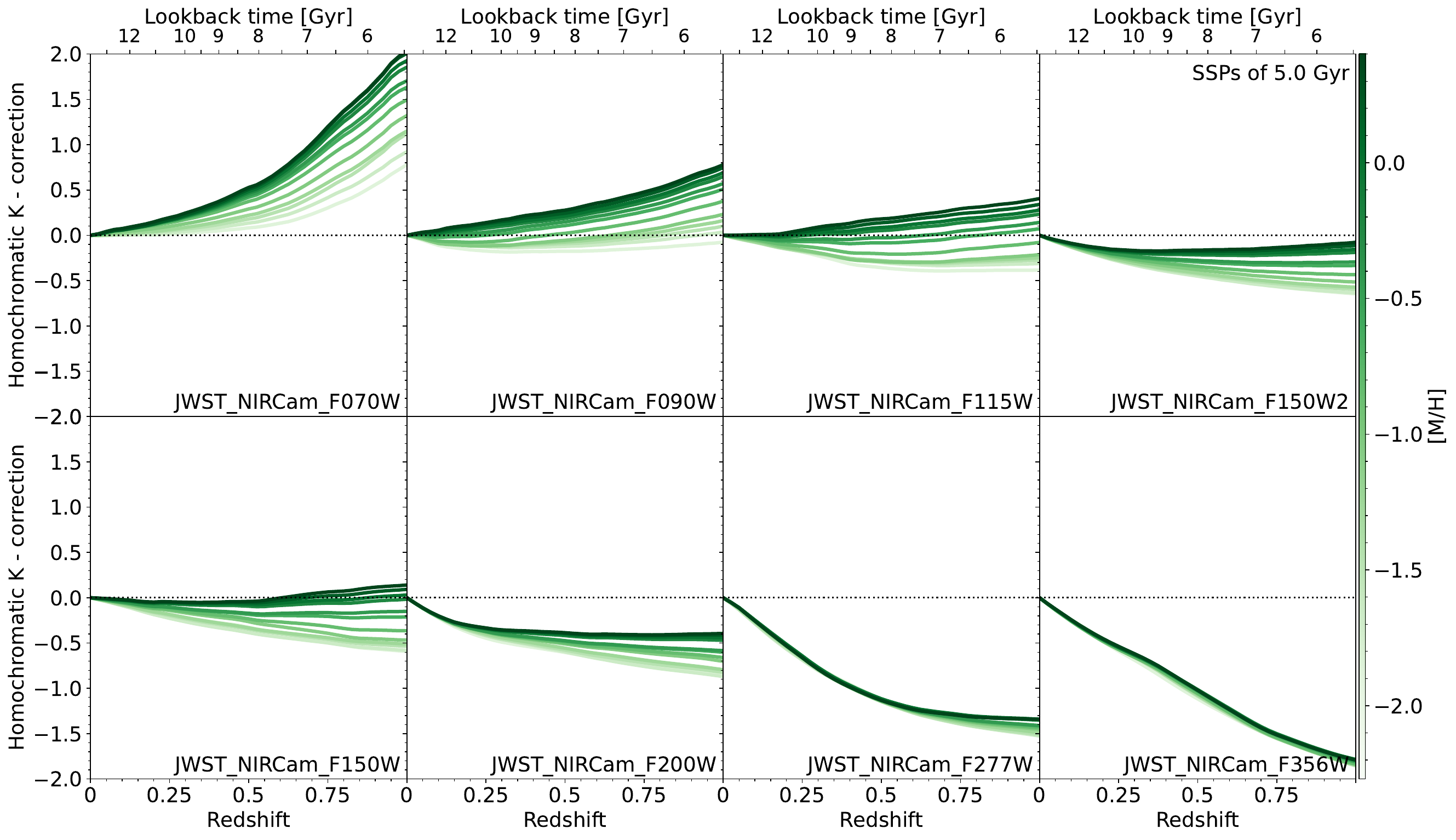}
    \caption{Homochromatic \K-corrections for the \jwst/NIRCam filters as a function of redshift for SSPs of $5~\gyr$ emitting at different redshifts. The solid lines correspond to different metallicities as indicated in the colourbar, and the black dotted line marks where no correction is needed (\K=$0$).} \label{fig:kcorr_homo_jwst}
\end{figure*}

%Homochromatic \K-values describe the correction needed to account for the effect of redshift within the same wavelength range. Although in principle they can be calculated for any filter, here we focus on the \K-corrections needed for the \jwst/NIRCam filters. These filters are allowing us to observe star cluster populations at intermediate redshifts for the first time \citep[e.g.][]{lee+2022,faisst+2022,harris_reina-campos2023}, and thus transforming them into a common frame is critical to be able to compare them.

We show homochromatic \K-corrections needed for a SSP of $5~\gyr$ observed with the \jwst filters in Fig.~\ref{fig:kcorr_homo_jwst}. For the bluest filter ($F070W$) and regardless of the metallicity of the SSP, the \K-corrections are always positive; that is the observed SED is dimmer than its intrinsic value, and thus require a brighter absolute magnitude.  The \K-values for the filters $F090W$, $F115W$, $F150W$, $F150W2$ and $F200W$ are well contained within the $[-1,1]$ magnitude range. Because the peak of the SED is much more prominent in SSPs of lower metallicities, the \K-values at low metallicities tend to show more variation over redshift and tend to be more negative. In the filters $F090W$, $F115W$ and $F150W$, the curves for some of the models lie near a null \K$=0$ until the peak of their SED finally shifts past the filter bandpass. 

For the filters of the \jwst/NIRCam long wavelength channel, $F277W$ and $F356W$, two effects are seen. Firstly, the variation of the SED with metallicity is so small that there is not much difference between their \K-values (see top panel in Fig.~\ref{fig:seds_mh_ages}). Secondly, the \K-values are quite large, \K$\geq -1.5$. Because the intrinsic luminosities are low in these bands, any radiation coming from around the peak and redshifted results in a large change of brightness, and thus, the absolute magnitudes have to be dimmed.

%\begin{figure*}
%	\includegraphics[width=\hsize]{imgs/homo_kcorr_hst_ssp_5gyr_mh.pdf}
%    \caption{Homochromatic K-corrections to be applied to SSPs of $5~\gyr$ and different metallicities calculated for different filters as a function of redshift of the source.} \label{fig:kcorr_homo_hst}
%\end{figure*}

\subsection{Heterochromatic \K-corrections}\label{sub:heterochromatic-ks}

\begin{figure*}
	\includegraphics[width=\hsize]{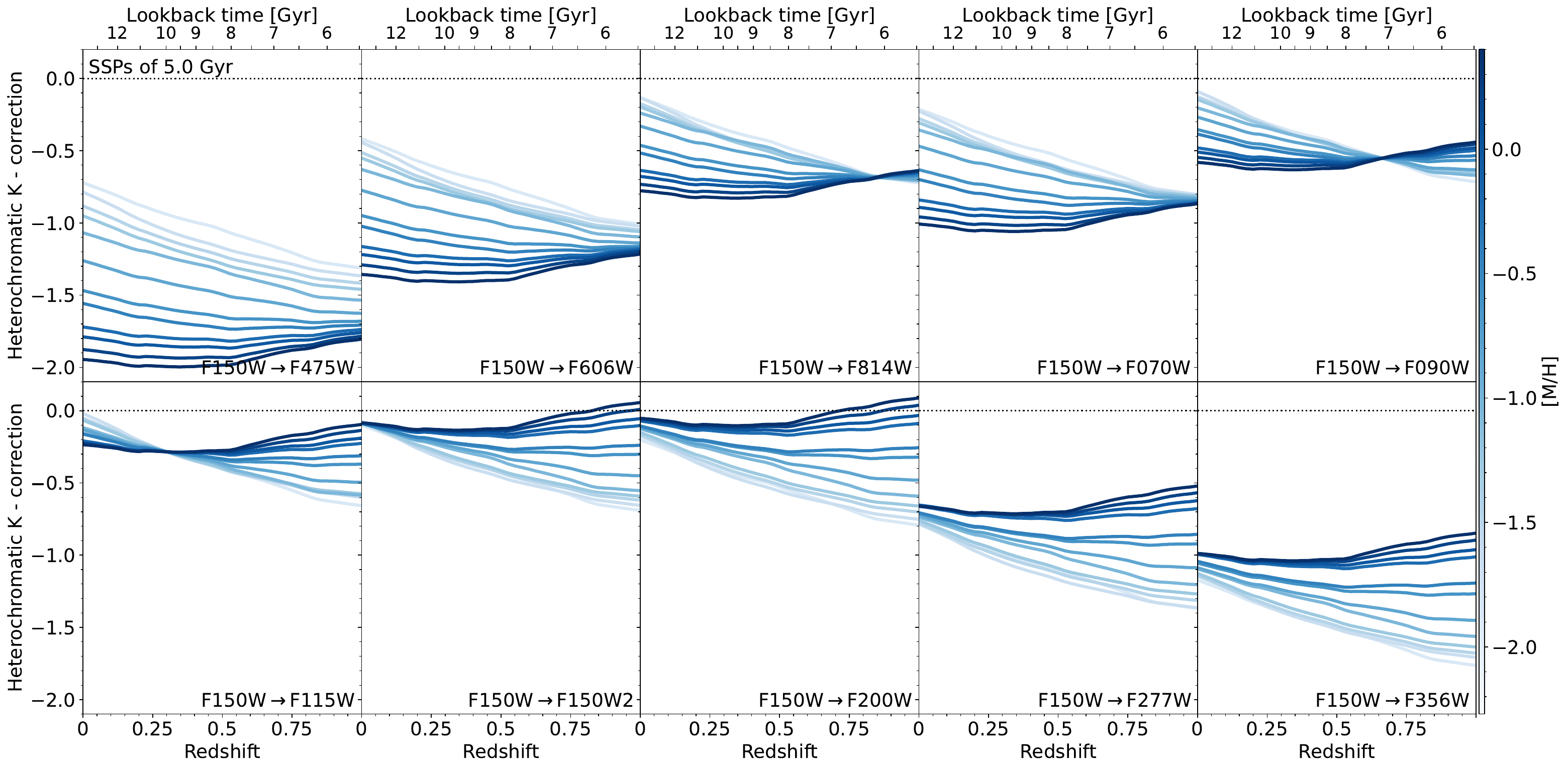}
    \caption{Heterochromatic \K-corrections as a function of redshift for SSPs of $5~\gyr$ emitting at different redshifts. Each panel shows the corrections calculated from the JWST/NIRCam $F150W$ filter to all the other filters in our set. The solid lines correspond to different metallicities as indicated in the colourbar, and the black dotted line marks where no correction is needed (\K=$0$).} \label{fig:kcorr_hetero}
\end{figure*}

In the case of heterochromatic \K-corrections, the transformation is performed from the observed filter towards any arbitrary filter. Thus, these corrections can allow comparisons between observations of star cluster populations in the \jwst filters and those in the \hst filter set.

We use the \jwst/NIRCam filter $F150W$ as our fiducial starting point for the observed wavelength range, and compute the corrections needed to transform it to any of the other filters in our set. We show the resulting \K-corrections for a SSP of $5~\gyr$ emitting as a function of redshift in Fig.~\ref{fig:kcorr_hetero}. Regardless of the metallicity of the SSP, the heterochromatic corrections are always negative. A trend is visible when correcting towards different parts of the spectrum: the corrections to the blue filters are negative and large (\K $\leq -1~\rm mags$), become smaller and closer to zero when transforming towards wavelengths $\lambda\sim 0.9$--$2~\mu$m, and increase again when correcting to the redder part of the spectrum. 
This trend results from the different shapes between the observed and the intrinsic SEDs, and can be seen from the green arrow in the right-hand panel of Fig.~\ref{fig:kcorr_pedagogical} tracing the intrinsic emitted curve.  

The correction towards filters redder than $F070W$ and bluer than $F200W$ leads to \K-values within $[0,-1]~\rm mags$. An interesting effect occurs in the transformations into the filters $F814W$, $F090W$, and $F115W$: at a particular redshift, the correction is the same regardless of the metallicity of the SSP. This is caused by the wavelength range of the target filter roughly corresponding to the wavelength range from which the photons were originally emitted. In the case of $F115W$, its midpoint wavelength at a redshift of $z=0.25$ is $11500~$\AA$~\times(1+0.25) = 14375~$\AA, very close to the midpoint wavelength $15000~$\AA~of the $F150W$ filter. At this particular lookback time, the luminosity terms in Eq.~(\ref{eq:kcorr-luminosity}) cancel out and the \K-correction is equal for all the SSPs. 

\section{Summary and Discussion}\label{sec:summary}

We present the photometric \K-corrections versus redshift for model star clusters that cover a wide range of ages and metallicities, for several broadband filters on the \hst/ACS and the \jwst/NIRCam cameras. Since star clusters are well described by simple stellar populations of a single age and metallicity, we use the spectral energy distributions for them from the E-MILES stellar library. 

The main effect of observing objects backwards in time is that their SEDs are attenuated and shifted towards redder wavelengths by the redshift. The \K-correction characterizes how much brighter or dimmer the observed SED is relative to the intrinsic luminosity of the object (Fig.~\ref{fig:kcorr_pedagogical}). In this work, we adapt the formalism developed for galaxy studies to describe the corrections needed for observations star clusters as a function of redshift. We have calculated these corrections both within the same wavelength range (i.e.~homochromatic corrections, Sect.~\ref{sub:homochromatic-ks}) and across filters (i.e.~heterochromatic corrections, Sect.~\ref{sub:heterochromatic-ks}).   

Observing remote star clusters has an interesting limitation: they have a well defined maximum observable age that decreases with increasing redshift. 
%Thus, a $\sim 13~\gyr$ old globular cluster at $z=0$ must have been only $2.61~\gyr$ old at $z=2.61$ (see Fig.~\ref{fig:kcorr_oldest}). 
Despite the attenuation due to redshift, the displacement of the SED peak into redder wavelengths and the increase in luminosity from the younger populations implies that high-$z$ star clusters remain bright in the infrared and should be detectable by \jwst up to $z\sim 1$.

All the \K-corrections are publicly available in a Zenodo repository, and we have developed an interactive webtool called RESCUER to generate the \K-values for any user-defined combination of cluster properties (App.~\ref{app:webtool}).

\section*{Acknowledgements}
The authors thank Laura Parker and David Hogg for productive comments.  MRC gratefully acknowledges the Canadian Institute for Theoretical Astrophysics (CITA) National Fellowship for partial support.  This work was supported by the Natural Sciences and Engineering Research Council of Canada (NSERC). 

\emph{Software:} E-MILES \citep{rock+2016} and BaSTI \citep{pietrinferni+2021}. This work has also made use of the following \code{Python} packages: \code{astropy}\footnote{\href{https://www.astropy.org}{https://www.astropy.org}} \citep{astropy:2013,astropy:2018,astropy:2022}, \code{h5py} \citep{h5py_allversions}, \code{Jupyter Notebooks} \citep{Kluyver16}, \code{Numpy} \citep{Harris20}, \code{Pandas} \citep{McKinney10, pandas_allversions},  \code{streamlit}\footnote{\href{https://streamlit.io}{https://streamlit.io}}, \code{Zenodo}\footnote{\href{https://zenodo.org}{https://zenodo.org}}, and all figures have been produced with the library \code{Matplotlib} \citep{Hunter07}.

%%%%%%%%%%%%%%%%%%%%%%%%%%%%%%%%%%%%%%%%%%%%%%%%%%
\section*{Data Availability}

The spectral models are publicly available in the E-MILES website: \href{http://research.iac.es/proyecto/miles/pages/ssp-models.php}{http://research.iac.es/proyecto/miles/pages/ssp-models.php}. 

The RESCUER interactive webtool is hosted by Streamlit in  \href{https://rescuer.streamlit.app/}{https://rescuer.streamlit.app/}, and we have deposited the tables with all the \K-corrections derived in this work in a Zenodo repository with DOI: 10.5281/zenodo.8387817. 


\bibliographystyle{mnras}
\bibliography{mybib} % if your bibtex file is called example.bib

%%%%%%%%%%%%%%%%%%%%%%%%%%%%%%%%%%%%%%%%%%%%%%%%%%

%%%%%%%%%%%%%%%%% APPENDICES %%%%%%%%%%%%%%%%%%%%%

\appendix

\section{\K-corrections for different IMFs and isochrones}\label{app:diff-seds}

Here we briefly explore the effect of the assumed stellar isochrone and IMF on the \K-corrections for a SSP of $\sim5~\gyr$ and $\mh\sim-1$ from the E-MILES stellar library. Changing the IMF to a \citet{kroupa2001} IMF leads to small changes in the intrinsic luminosity ($\ll 10^{28}~\rm ergs~s^{-1}~$\AA$^{-1}~\rm M_{\odot}^{-1}$), whereas changing the stellar isochrones to the Padova models \citep{girardi2000} produces SEDs that are $\sim 10^{28}~\rm ergs~s^{-1}~$\AA$^{-1}~\rm M_{\odot}^{-1}$ brighter. When propagating those changes to the \K-corrections, we find that for the \jwst SWC filter $F150W$ the homochromatic correction is $0.02~\rm mags$ higher at $z=0.5$ and $0.12~\rm mags$ higher at $z=0.1$ when assuming the Padova models relative to the BaSTI isochrones. Similar offsets are found in the heterochromatic \K-correction between the \jwst SWC filters $F150W$ and $F115W$.

\begin{figure}
 	\includegraphics[width=\hsize]{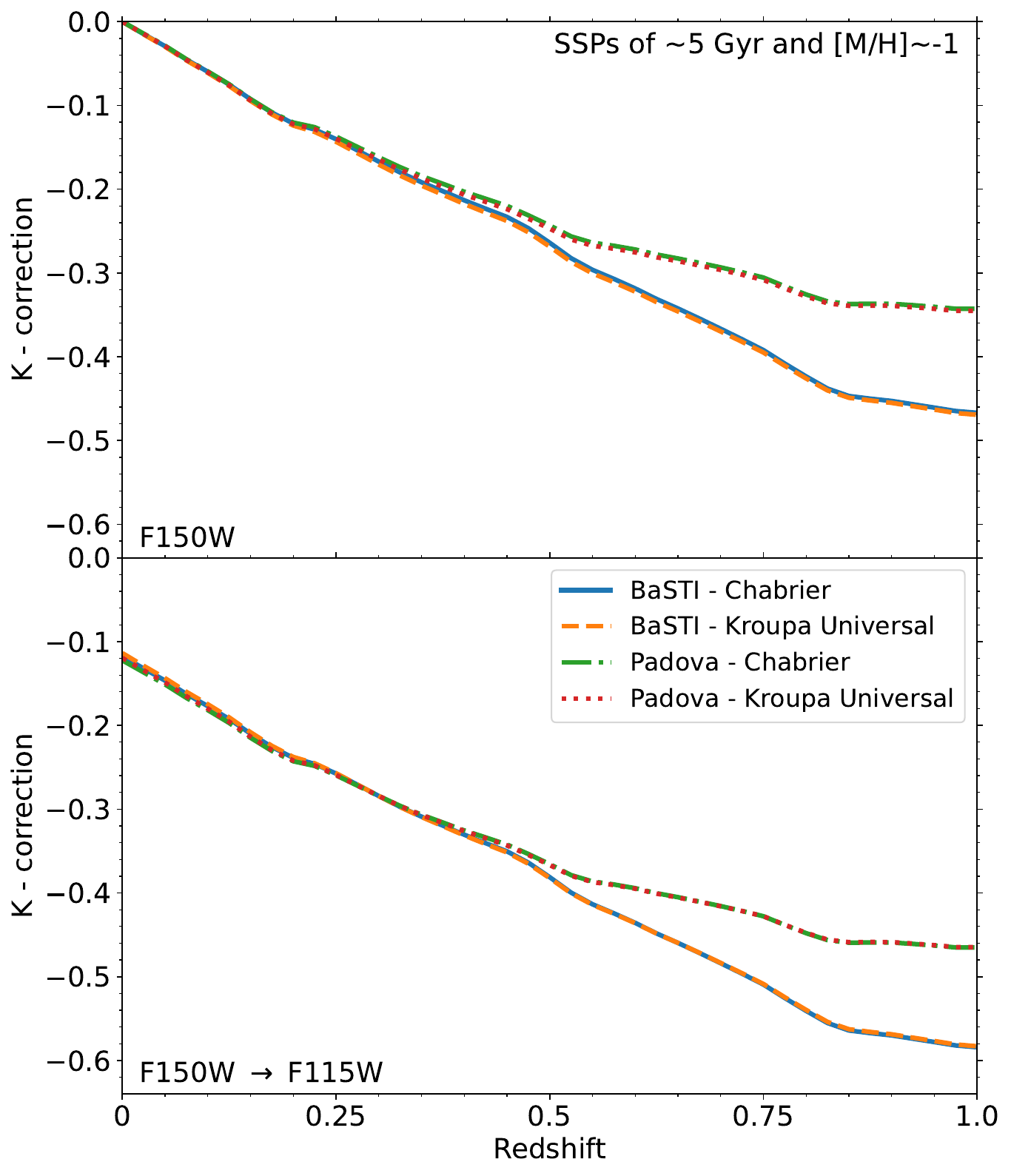}
    \caption{Comparison of the predicted luminosity for a SSP of $\sim 5~\gyr$ old and $\mh\sim -1$ assuming different isochrones and stellar IMFs as indicated in the legend.} \label{fig:sed_alternative}
\end{figure}

\section{RESCUER: \K-corrections via a webtool}\label{app:webtool}

To facilitate the calculation of the photometric \K-corrections for star clusters (i.e.~SSPs of a given age and metallicity), we have developed a webtool that allows users to do so interactively.  The code is called RESCUER (for REdshifted Star ClUstERs). The current version of the webtool calculates the cosmological \K-corrections for SSPs from the E-MILES stellar library assuming BaSTI isochrones and a Chabrier IMF for the same set of \hst and \jwst filters presented here. The webtool can be found in \href{https://rescuer.streamlit.app}{https://rescuer.streamlit.app}, and its source code is hosted in the public GitHub repository \href{https://github.com/mreinacampos/rescuer/}{https://github.com/mreinacampos/rescuer/}.

%%%%%%%%%%%%%%%%%%%%%%%%%%%%%%%%%%%%%%%%%%%%%%%%%%

% Don't change these lines
\bsp	% typesetting comment
\label{lastpage}
\end{document}